\documentclass[twocolumn,english,aps,prl]{revtex4}
\usepackage{amssymb}
\usepackage[latin1]{inputenc}
\usepackage{amsmath}
\usepackage{graphicx}
\usepackage{sistyle}
\usepackage{color}
\makeatletter

\usepackage{subfigure}
\begin{document}

\title{Oscillatory transient regime in the forced dynamics of a spin torque nano-oscillator}

\author{Yan Zhou$^1$}
\email{zhouyan@kth.se}

\author{Vasil Tiberkevich$^{2}$}

\author{Ezio Iacocca$^1$}

\author{Andrei Slavin$^{2}$}

\author{Johan \AA kerman$^{1,3}$}
\email{akerman1@kth.se}

\affiliation{
$^1$~Department of Microelectronics and Applied Physics, Royal Institute of Technology, Electrum 229, 164 40 Kista, Sweden
\\
$^2$~Department of Physics, Oakland University, Rochester, MI 48309 USA
\\
$^3$~Physics Department, G\"oteborg University, 412 96 G\"oteborg, Sweden
}

\begin{abstract}
We demonstrate that the transient non-autonomous dynamics of a spin torque nano-oscillator (STNO) under a radio-frequency (rf) driving signal is qualitatively different from the dynamics described by the Adler model. If the external rf current
$I_{rf}$ is larger than a certain critical value $I_{cr}$
(determined by the STNO bias current and damping) strong
oscillations of the STNO power and phase develop in the transient
regime. The frequency of these oscillations increases with $I_{rf}$
as $\propto\sqrt{I_{rf} - I_{cr}}$ and can reach several GHz,
whereas the damping rate of the oscillations is almost independent
of $I_{rf}$. This oscillatory transient dynamics is caused by the
strong STNO nonlinearity and should be
taken into account in most STNO  rf applications.
\end{abstract}

\maketitle
Interest in spin torque nano-oscillators (STNOs) \cite{Katine2000,
Houssameddine2007, Katine2008, Ralph2009} is rapidly growing among researchers as these nano-scale auto-oscillating systems have fascinating nonlinear and non-trivial properties \cite{Slavin2009}. STNOs are also interesting for rf applications due
to an attractive combination of their properties, such as a wide
range of generated frequencies \cite{Bonetti2009}, fast modulation
rates \cite{Pufall2005APL}, and easy integration into the modern
on-chip nano-electronic circuits.

Recently, several groups have performed studies of the forced dynamics of an STNO under the influence of external microwave current \cite{Rippard2005, Li2006PRB, Zhou2007,
Georges2008a}. As with other types of auto-oscillatory systems, such
dynamics results in a synchronization, or injection locking, of STNO
oscillations to the external signal. The classical theory of injection locking developed by Adler in 1946 \cite{Adler1946} is typically used to analyze such experiments on STNOs. By treating the oscillator as an active element
coupled to a resonant circuit, Adler obtained a simple dynamical
equation for the phase difference $\psi$ between the
auto-oscillation and the injected driving signal:
\begin{equation}
\label{eq:Adler}
    \frac{d\psi}{dt} = -\Delta\omega - F\sin(\psi)
\ .\end{equation} Here $\Delta\omega = \omega_e - \omega_0$ is the
mismatch between the frequency of the injected signal $\omega_e$ and
the frequency of free auto-oscillations $\omega_0$, and $F$ is
proportional to the amplitude of the injected signal. Despite its simplicity, Eq.~(\ref{eq:Adler}) not only describes phase-locking phenomena in a variety of different physical systems, but accounts for all major characteristics of the phase-locking process. In
particular, Eq.~(\ref{eq:Adler}) predicts the frequency interval of
phase-locking $|\Delta\omega| < F$ and stationary phase relation
$\psi_A = - \arcsin(\Delta\omega/F)$ between the driving signal and the
locked oscillation. Additionally, it follows from
Eq.~(\ref{eq:Adler}) that the phase $\psi$ approaches its locked
value $\psi_A$ \emph{monotonically} as an exponential
with a time constant $\tau_A = 1/(F\cos\psi_A)$ inversely
proportional to the driving signal $F$.

In this Letter, we show that for sufficiently strong injected
microwave current $I_{rf}$, or modulation depth $\mu = I_{rf}/I_{dc}$
(where $I_{dc}$ is the dc bias current), Adler's model breaks down
and does not give an adequate description of the phase-locking of an
STNO. The most striking discrepancies are \emph{i}) pronounced
transient \emph{oscillations} of the STNO phase difference $\psi$
during its approach to phase locking, and \emph{ii}) a
synchronization time $\tau_s$ which is \emph{independent} of the
driving amplitude $I_{rf}$. We will show that these qualitative features
are due to the strong nonlinearity of the STNO. Additionally, we find the critical modulation depth, $\mu_{cr}$, separating Adlerian and non-Adlerian
dynamics is a surprisingly small quantity. We therefore conclude that phase locking is almost always non-Adlerian in practical STNO devices.

The STNO dynamics is studied numerically within a standard
macrospin approximation \cite{sunjz2000, Zhou2007, xiaoj2005,
Zhou2008TiltSTOAPL}. The normalized (unit-length) magnetization
vector ${\bf m}$ of the STNO free layer obeys the
Landau-Lifshitz-Gilbert-Slonczewski (LLGS) equation
\cite{Slonczewski1996, xiaoj2005}:
\begin{equation}
\label{eq:LLGS}
    \frac{d{\bf m}}{dt} = -|\gamma|{\bf m}\times{{\bf H}_{eff}} + \alpha{\bf m}\times\frac{d{\bf m}}{dt} + |\gamma|\alpha_J{\bf m}\times({\bf m}\times{\bf p})
\ .\end{equation} Here $\gamma = - 1.76 \times 10^{11}$~Hz/T is the
gyromagnetic ratio, ${\bf{H}}_{eff} = H_a{\bf z} - M_s({\bf
m}\cdot{\bf z}){\bf z}$ is the effective magnetic field (where
$\mu_0H_a = 1.5$~T is the external magnetic field applied, in the
studied case, along the normal ${\bf z}$ to the free STNO layer and
the second term is the demagnetization field with $\mu_0M_s = 0.8$~T
being the free layer saturation magnetization), $\alpha = 0.01$ is
the Gilbert damping constant. $\alpha_J$ is the spin torque
magnitude defined as $\alpha_J = \hbar \eta_0 I/(2 \mu_0 M_s e
V)$, where $\hbar$ is the Planck constant, $\eta_0 = 0.35$ is the
dimensionless spin torque efficiency, $I$ is the applied current,
$\mu_0$ is the free space permeability, $e$ is the fundamental
electric charge, and $V = 3\times 10^4$~nm$^3$ is the volume of the
free layer. The unit vector ${\bf p} = \cos(\gamma_0){\bf z} +
\sin(\gamma_0){\bf x}$ in Eq.~(\ref{eq:LLGS}), which defines the
direction of current spin polarization, coincides with the
magnetization direction of the fixed STNO layer. In our simulations
we used a tilt angle $\gamma_0 = 60^\circ$.

For a constant dc current $I = I_{dc}$ the last term in
Eq.~(\ref{eq:LLGS}) describes an effective negative damping that
compensates the natural positive magnetic damping (second term on
the right hand side of Eq.~(\ref{eq:LLGS})). When the bias current
$I_{dc}$ exceeds a certain threshold value $I_{th}$ (2.32 mA in our
case), a stable precession of the magnetization vector $\bf m$
develops in the STNO  and the free-running frequency $\omega_0$ of
this precession depends on both ${\bf H}_{eff}$ and $I_{dc}$. In our
simulations we used $I_{dc} = 3$~mA (supercriticality parameter
$\zeta = I_{dc}/I_{th} = 1.29$), in which case the free STNO
frequency was $\omega_0/(2\pi) = 25.3$~GHz.

In the forced regime, when in addition to the bias current
$I_{dc}$ the STNO is driven by an injected microwave current
$I_{rf}$, i.e. $I(t) = I_{dc} + I_{rf}\sin(\omega_e t)$ with
$\omega_e$ close to $\omega_0$, the STNO may phase-lock to $I_{rf}$.
In the phase-locked regime the generated STNO frequency becomes
exactly equal to $\omega_e$ and a fixed (independent of the initial
conditions) phase difference $\psi_0$ develops between $I_{rf}$ and
the STNO oscillation.

\begin{figure}[t!]
\includegraphics[scale=0.7, clip=true, viewport=4.8in 2.2in 15in 5.8in]{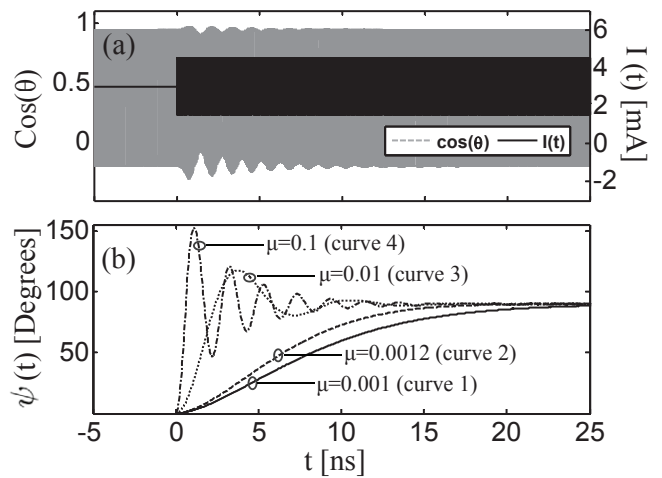}
\caption{\label{f:basics} (a) Time dependence of the STNO signal
$\cos(\theta(t)) = {\bf m}(t)\cdot{\bf p}$ and (b) transient
behavior of the phase difference between the STNO signal and
$I_{rf}$ after $I_{rf}$ was switched on at $t = 0$. $\mu =
I_{rf}/I_{dc} = 0.5$ in (a) and $\mu = 0.001$, $0.0012$, $0.01$, and
$0.1$ for curves 1--4, respectively, in (b). }
\end{figure}

The results of the numerical simulations describing the STNO
approach to phase-locking are shown in Fig.~\ref{f:basics} for
various $I_{rf}$ amplitudes. The STNO was first prepared in a
free-running state ($I_{rf} = 0$) for 50~ns to achieve a stable
free-running regime. At $t = 0$ the microwave current $I_{rf}$ was
switched on with $\omega_e = \omega_0$. Since $\omega_e$ and
$\omega_0$ coincide, the phase locking manifests itself only by
establishing a fixed phase relations between these oscillations.
Fig.~\ref{f:basics}(a) shows the time dependence of $\cos(\theta(t))
= {\bf m}(t)\cdot{\bf p}$ (which is proportional to the STNO output
signal) for the modulation depth $\mu = I_{rf}/I_{dc} = 0.5$. One
can clearly see a transient beating of the envelope of the STNO
signal, which indicates an \emph{oscillatory} approach to the phase
locking. This oscillatory approach is shown explicitly in
Fig.~\ref{f:basics}(b), where we plot the time dependence of the
phase difference $\psi(t) = \phi(t) - \omega_e t$ between the STNO
phase $\phi(t) = \arctan(m_y/m_x)$ and the phase of the external
signal $\omega_e t$ for several values of the modulation depth
$\mu$.

We first note that the stationary value of the phase difference is
substantially different from zero ($\psi_0 \approx 90^\circ$) in
contrast with what one would expect from Eq.~(\ref{eq:Adler}) for
$\Delta\omega = \omega_e - \omega_0 = 0$. This significant
\emph{intrinsic} phase shift \cite{Zhou2007,Zhou2008} is caused by
the strong nonlinearity of the STNO generation frequency
\cite{Slavin2009}. One can also see from Fig.~\ref{f:basics}(b) that
the transient dependence $\psi(t)$ is monotonic only for extremely
small values of $\mu$, whereas for all the reasonable modulation
depths strong oscillations of phase  develop in the transient
regime. The critical modulation depth, separating regions of the
monotonic (Adlerian) and non-monotonic (non-Adlerian) dynamics is as
small as $\mu_{cr} = 0.0012$ (curve 2 in Fig.~\ref{f:basics}(b)).

Another striking feature of the non-Adlerian transient regime is the very weak dependence of the synchronization time (the time
needed to achieve the locked state) on the normalized driving
amplitude $I_{rf}$. For example, the envelopes of the curves 3 and 4 in
Fig.~\ref{f:basics}(b) have essentially the same time constant
$\tau_s$, whereas the classical Adler's model Eq.~(\ref{eq:Adler})
predicts that the time constant in these two cases should differ by
a factor of 10.

\begin{figure}[t!]
\includegraphics[scale=0.7, clip=true, viewport=4.9in 2.2in 15in 5.8in]{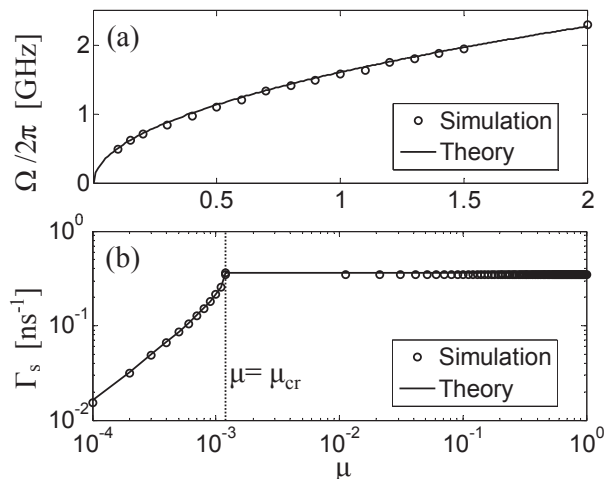}
\caption{\label{f:dependence} Dependence on $\mu$ of (a) the
frequency of the transient STNO phase oscillations $\Omega$, and (b)
its decay constant $\Gamma_s$. Points -- results of numerical
simulations of Eq.~(\ref{eq:LLGS}); solid lines -- analytical
expressions Eq.~(\ref{eq:omega}) and Eq.~(\ref{eq:gamma}). The
vertical line indicates the critical modulation depth $\mu_{cr}$. }
\end{figure}

The non-Adlerian STNO dynamics are further illustrated in
Fig.~\ref{f:dependence}, where we show the dependence of the
frequency of the transient phase oscillations $\Omega$ on $\mu$ (panel
(a)) and the decay constant $\Gamma_s = 1/\tau_s$ of these
oscillations (panel (b)). One can see that $\Omega$ increases with
$\mu$ and reaches GHz values for accessible modulation depths
$\mu \sim 0.5$. The decay rate $\Gamma_s$ increases approximately
linearly with $\mu$ (following the Adler's model Eq.~(\ref{eq:Adler}))
only for $\mu < \mu_{cr}$, whereas in the non-Adlerian region $\mu >
\mu_{cr}$ it remains virtually constant.

The results shown in Fig.~\ref{f:basics} and Fig.~\ref{f:dependence}
clearly demonstrate that the  STNO  behavior in the transient
phase-locking process can not be described by a classical Adler's
model Eq.~(\ref{eq:Adler}). Although the transient oscillations may not be
visible in locking experiments performed in a stationary
regime, they may significantly influence the STNO operation in
applications where ultra-fast transitional dynamics are important.
For instance, Fig.~\ref{f:dependence}(b) shows that the
synchronization time constant $\tau_s = 1/\Gamma_s$ cannot be
reduced below $\approx 3$~ns by increasing the amplitude of the
injected current, which may limit the operational speed of STNOs as
ultra-fast signal modulators.

To understand the origin of the non-Adlerian dynamics discussed above we note that STNOs, in contrast to the majority of conventional auto-oscillators, demonstrate a strong dependence of the generated frequency, $\omega(p)$, on the generated power $p$ \cite{Slavin2009}.
%
As a result, even small power fluctuations
$\delta p = p - p_0$ from the free-running power $p_0$ may result in
significant deviations of the generated frequency $\omega(p) -
\omega(p_0) \approx N\delta p$ (here $N = d\omega(p)/dp$ is the
nonlinear frequency shift coefficient). Equations describing STNO phase locking that also take into account power fluctuations have been derived in \cite{Slavin2009} and can be written as:
\begin{subequations}\label{eq:system2}
\begin{eqnarray}
\label{eq:psi2}
    \frac{d\psi}{dt} &=& -\Delta\omega -F\sin\psi +N\delta p\,,\\
\label{eq:x2}
    \frac{d\delta p}{dt} &=& -2\Gamma_p\delta p + 2p_0F\cos\psi\ .
\end{eqnarray}
\end{subequations}
Here $\psi(t) = \phi(t) - \omega_e t$ is the phase difference
between the STNO signal and the external signal, $F$ is the
normalized external signal amplitude, and $\Gamma_p$ is the damping
rate of power fluctuations. For the geometry studied here, all the parameters of Eqs.~(\ref{eq:system2}) can be calculated analytically
\cite{Slavin2009} to give: $p_0 = (\zeta - 1)\omega_H/(2\omega_M)$,
$N = 2\omega_M$, $\Gamma_p = \alpha\omega_H(\zeta - 1)$, $F =
\mu\cdot\alpha\omega_H\tan(\gamma_0)/(4\sqrt{p_0})$, where $\zeta =
I_{dc}/I_{th}$ is the supercriticality parameter, $\omega_H =
|\gamma|(H_a - M_s)$ is the ferromagnetic resonance (FMR) frequency,
and $\omega_M = |\gamma|M_s$, and these expressions are valid for
moderate supercriticalities $\zeta \leq 1.5$. For a linear ($N = 0$)
oscillator Eq.~(\ref{eq:psi2}) coincides with the Adler's model
Eq.~(\ref{eq:Adler}).

The stable stationary (phase-locked) solution of Eqs.~(\ref{eq:system2}) has the form
\begin{subequations}\label{eq:statsol}
\begin{eqnarray}
\label{eq:psistat}
    \psi_0 &=& \arctan(\nu)-\arcsin\left({\Delta\omega}/{\Delta\omega_0}\right)\,,\\
\label{eq:xstat}
    \delta p_0 &=& p_0\frac{\nu\Delta\omega+\sqrt{\Delta\omega_0^2-\Delta\omega^2}}{(1+\nu^2)\Gamma_p}\,,
\end{eqnarray}
\end{subequations}
where $\nu = Np_0/\Gamma_p$ is the dimensionless nonlinearity
parameter (in our case $\nu \approx 1/\alpha = 100$) and
$\Delta\omega_0 = \sqrt{1 + \nu^2}F$ is nonlinearity-enhanced
frequency interval of phase-locking \cite{Slavin2005}. The first
term in Eq.~(\ref{eq:psistat}) describes the above mentioned
intrinsic phase shift of a strongly nonlinear STNO.

By linearizing Eqs.~(\ref{eq:system2}) near the solution,
 Eqs.~(\ref{eq:statsol}), one can find the decay rate, $\lambda$, of
phase and power deviations from the stationary phase-locked state:
\begin{align}
\label{eq:solution}
    \lambda = &\Gamma_p + \frac{1}{2}F\cos\psi_0\nonumber\\
&\pm\sqrt{\left(\Gamma_p - \frac{1}{2}F\cos\psi_0\right)^2 -
2\nu\Gamma_p F\sin\psi_0} \ .\end{align}

For a quasi-linear ($\nu = 0$), or Adlerian, auto-oscillator,
Eq.~(\ref{eq:solution}) gives $\lambda_1 = 2\Gamma_p$ and $\lambda_2
= F\cos\psi_0$. The decay rate $\lambda_1$ describes the damping
rate of the power deviations $\delta p$, whereas the rate
$\lambda_2$ corresponds to the decay of the pure phase deviations
$\psi - \psi_0$. The synchronization time $\tau = 1/\lambda$
quantifies the overall time needed to reach a phase-locked state.
For realistic parameters $\Gamma_p \gg F$ so the locking time of an
Adlerian oscillator is determined by the transient phase dynamics
and is given by $\tau_A = 1/(F\cos\psi_0) \propto 1/F$.

To analyze the case of a strongly nonlinear ($|\nu| \gg 1$) STNO, we
note that in this case one can neglect $F\cos\psi_0$ and simplify
Eq.~(\ref{eq:solution}) to
\begin{equation}\label{eq:solution2}
    \lambda = \Gamma_p\left(1 \pm \sqrt{1 - {F}/{F_{cr}}}\right) = \Gamma_p\left(1 \pm \sqrt{1 - {\mu}/{\mu_{cr}}}\right)
\,,\end{equation} where the critical signal amplitude is  $F_{cr} =
\Gamma_p/(2\nu\sin\psi_0)$ or, in terms of the modulation depth of
the above considered  STNO with a perpendicularly magnetized free
layer can be expressed as
\begin{equation}
\label{eq:mucr}
    \mu_{cr} \approx \frac{\alpha}{\tan\gamma_0}(\zeta-1)^{3/2}\sqrt{\frac{2\omega_H}{\omega_M}} = 0.0012
\ .\end{equation}

For modulation depths $\mu > \mu_{cr}$ the decay rates $\lambda$
become complex (see Eq.~(\ref{eq:solution2})), and describe an
oscillatory approach to the phase-locked state. The frequency of
these transient oscillations is given by
\begin{equation}\label{eq:omega}
    \Omega = \Gamma_p\sqrt{\mu/\mu_{cr} - 1}
\end{equation}
and is shown as a solid line in Fig.~\ref{f:dependence}(a) for the
calculated value $\Gamma_p = 0.36$~ns$^{-1}$. One can see an
excellent agreement between the numerical and analytical results.

The decay constant $\Gamma_s$ of the phase oscillation is given by
the smallest of the $\lambda$'s in Eq.~(\ref{eq:solution2}) when
both of them are real (Adlerian regime), and is equal to the real
part of the $\lambda$'s, when they are complex (non-Adlerian
regime):
\begin{equation}\label{eq:gamma}
    \Gamma_s = \left\{\begin{array}{ll}
        \Gamma_p\left(1 - \sqrt{1 - \mu/\mu_{cr}}\right)\,,\qquad & \mu < \mu_{cr}
\\
        \Gamma_p\,,\qquad & \mu > \mu_{cr}
    \end{array}\right.
\ .\end{equation} The dependence of Eq.(\ref{eq:gamma}) is shown as a
solid line in Fig.~\ref{f:dependence}(b), and the agreement between
the simulations and the analytical calculations is again remarkable.

Thus, our analysis explains the observed non-Adlerian behavior of
the transient  STNO phase-locking as a result of a nonlinear
coupling between the power and phase fluctuations, and provides
quantitative expressions for both the frequency Eq.~(\ref{eq:omega})
and damping rate Eq.~(\ref{eq:gamma}) of the transient phase
oscillations.


We would also like to stress that the critical modulation depth
$\mu_{cr}$, defining the boundary between the Adlerian and
non-Adlerian dynamics (see Eq.~(\ref{eq:mucr})), is generally a
\emph{small} quantity, since $\alpha \ll 1$. Consequently, the
critical microwave current $I_{cr} = \mu_{cr} I_{dc}$ is typically
of the order of \SI{1}{\mu A}, which is much smaller than the
typical injection currents used in experiments. In other words,
phase locking of STNOs almost always takes place in the non-Adlerian
regime.

\begin{figure}[t!]
\includegraphics[scale=0.66, clip=true, viewport=4.9in 1.56in 15in 5.96in]{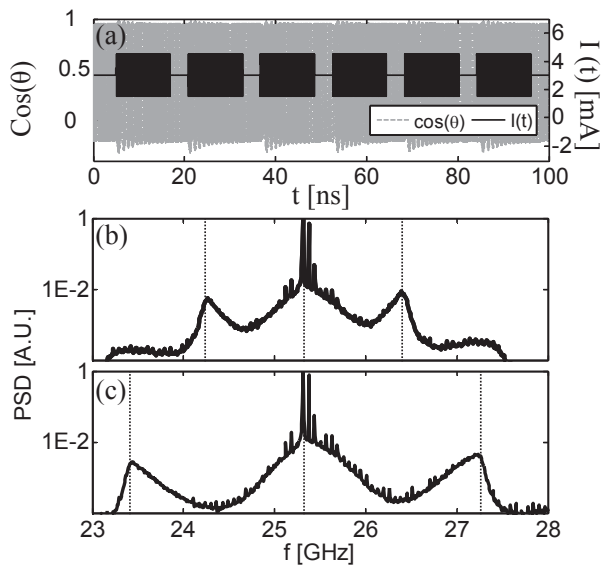}
\caption{\label{fig:PulsedRF} Phase-locking of an STNO to a pulsed
microwave signal with the repetition period of 16~ns. (a) Time
dependence of the STNO signal for $\mu = 0.5$. (b) Spectrum of the
STNO oscillations in (a). (c) Same for $\mu = 1.5$. Vertical lines
in (b) and (c) indicate the free-running frequency $\omega_0$ and
expected positions of the sidebands $\omega_0 \pm \Omega$. (Other
narrow sidebands are due to the direct frequency modulation at the
pulse repetition rate.)}
\end{figure}

We finally suggest a way to observe the transient phase oscillations of
an STNO experimentally. If the injected current is pulsed with the
repetition rate of the order of $1/\Gamma_p$, large sidebands at the
frequencies $\omega_0 \pm \Omega$ should appear in the spectrum of
the  STNO oscillations. Since the transient frequency $\Omega$ may
be significantly larger than the STNO generation linewidth, both the
position and shape of these sidebands can be measured
experimentally, providing important information about such intrinsic
STNO parameters as the nonlinear frequency shift $N$ and the damping
rate $\Gamma_p$ of power fluctuations. In Fig.~\ref{fig:PulsedRF} we show the results of numerical simulations of an STNO subjected to a \emph{pulsed} rf driving signal with a repetition period of 16~ns. Fig.~\ref{fig:PulsedRF}(a) shows the temporal profile of the
STNO signal for a modulation depth of $\mu = 0.5$ demonstrating
well-resolved intrinsic oscillations of the STNO power.
Fig.~\ref{fig:PulsedRF}(b) and (c) show the spectrum of STNO
oscillations for two values of the modulation depth $\mu$. One can
clearly see the sidebands caused by the intrinsic transient STNO
phase oscillations and their expected dependence on the modulation
depth.

In conclusion, we have shown that the transient forced
dynamics of an STNO for a sufficiently strong external signal cannot
be described by the classical Adler's model. The reason for this
non-Adlerian behavior is the strong nonlinearity of the STNO
generation frequency, which couples power and phase fluctuations. As
a result, strong phase oscillations in the GHz frequency range develop during the transient regime of phase-locking. The same
nonlinear mechanism determines the lower limit for the STNO
synchronization time (of the order of the characteristic decay
time $1/\Gamma_p$ of the STNO power fluctuations) and will, likely,
limit the maximum speed of the STNO frequency modulation.

\begin{acknowledgments}
We thank S. Bonetti for fruitful discussions. We gratefully
acknowledge financial support from The Swedish Foundation for
strategic Research (SSF), The Swedish Research Council (VR), the
G\"{o}ran Gustafsson Foundation, by the Contract no.
W56HZV-09-P-L564 from the U.S. Army TARDEC and RDECOM, by the Grant
no. ECCS-0653901 from the National Science Foundation of the USA.
Johan {\AA}kerman is a Royal Swedish Academy of Sciences Research
Fellow supported by a grant from the Knut and Alice Wallenberg
Foundation.
\end{acknowledgments}


\end{document}